\documentclass[preprint]{aastex}
\newcounter{address}
\newcommand{\equationname}{Equation}
\newcommand{\etal}{\emph{et al.}}
\newcommand{\eg}{\emph{e.g.}}
\newcommand{\ie}{i.e.}
\renewcommand{\mag}{\mathrm{mag}}
\renewcommand{\exp}[1]{\mathrm{e}^{#1}}
\newcommand{\true}[1]{{#1}_{\mathrm{true}}}
\newcommand{\obs}[1]{{#1}_{\mathrm{obs}}}
\newcommand{\DA}{D_{\mathrm{A}}}
\newcommand{\DL}{D_{\mathrm{L}}}
\newcommand{\DM}{DM}
\newcommand{\DV}{D_{\mathrm{V}}}
\newcommand{\DHub}{D_{\mathrm{H}}}
\newcommand{\drm}{{\mathrm{d}}}
\newcommand{\km}{{\mathrm{km}}}
\newcommand{\s}{{\mathrm{s}}}
\newcommand{\Mpc}{{\mathrm{Mpc}}}

\begin{document}
\setlength{\hbadness}{10000}

\title{Cosmic transparency:\\
       A test with the baryon acoustic feature and type Ia supernovae}
\author{Surhud More\altaffilmark{\ref{MPIA},\ref{email}},
  Jo Bovy\altaffilmark{\ref{CCPP}}, and
  David W. Hogg\altaffilmark{\ref{MPIA},\ref{CCPP}}}

\setcounter{address}{1}
\altaffiltext{\theaddress}{\stepcounter{address}\label{MPIA}
Max-Planck-Institut f\"ur Astronomie, K\"onigstuhl 17,
D-69117 Heidelberg, Germany}
\altaffiltext{\theaddress}{\stepcounter{address}\label{email} To whom
correspondence should be addressed: \texttt{more@mpia-hd.mpg.de}}
\altaffiltext{\theaddress}{\stepcounter{address}\label{CCPP} Center
for Cosmology and Particle Physics, Department of Physics, New York
University, 4 Washington Place, New York, NY 10003}


\begin{abstract}
  Conservation of the phase-space density of photons plus Lorentz
  invariance requires that the cosmological luminosity distance be
  larger than the angular diameter distance by a factor of $(1+z)^2$,
  where $z$ is the redshift.  Because this is a fundamental symmetry,
  this prediction---known sometimes as the ``Etherington relation'' or
  the ``Tolman test''---is independent of the world model, or even the
  assumptions of homogeneity and isotropy.  It depends, however, on
  Lorentz invariance and transparency.  Transparency can be affected
  by intergalactic dust or interactions between photons and the dark
  sector.  Baryon acoustic feature (BAF) and type Ia supernovae
  (SNeIa) measures of the expansion history are differently sensitive
  to the angular diameter and luminosity distances and can therefore
  be used in conjunction to limit cosmic transparency.  At the present
  day, the comparison only limits the change $\Delta\tau$ in the
  optical depth from redshift $0.20$ to $0.35$ at visible wavelengths
  to $\Delta\tau < 0.13$ at 95\% confidence.  In a model with a
  constant comoving number density $n$ of scatterers of constant
  proper cross-section $\sigma$, this limit implies $n\,\sigma<
  2\times10^{-4}\,h~\Mpc^{-1}$.  These limits depend weakly on
  cosmological world model. Assuming a concordance world model, the
  best-fit value of $\Delta\tau$ to current data is negative at the
  2$\sigma$ level. This could signal interesting new physics or could be
  the result of unidentified systematics in the BAF/SNeIa
  measurements. Within the next few years, the limits on transparency
  could extend to redshifts $z\approx2.5$ and improve to
  $n\,\sigma<1.1 \times10^{-5}\,h~\Mpc^{-1}$.  Cosmic variance will
  eventually limit the sensitivity of any test using the BAF
  at the $n\,\sigma\sim 4\times10^{-7}\,h~\Mpc^{-1}$
  level.  Comparison with other measures of the transparency is
  provided; no other measure in the visible is as free of
  astrophysical assumptions.
\end{abstract}



\section{Introduction}

The transparency of the universe is extremely good.  A typical
astronomical camera has a shutter whose thickness is measured in
microns; that shutter is far more opaque than the entire line of sight
to the majority of extragalactic sources, even at extremely high
redshifts, despite---in many cases---considerable column depths of
dark matter, plasma, gas, and dust.  There are,
however, very few quantitative measures of the transparency with
contemporary astronomical data.

There are several sources for photon attenuation that are clustered
with matter.  For example, as stars eject heavy elements, they also
eject photon-absorbing ash (called ``dust'').  The gas and plasma in
and around galaxies absorbs, scatters, and re-emits at longer
wavelengths some fraction of incident radiation.  More speculatively,
if the dark matter is an axion or axionlike particle, it will in
general have photon interactions, which can in principle produce
effective absorption of photons in regions of high dark matter density
and high magnetic fields
\citep{1983PhRvL..51.1415S,1988PhRvD..37.1237R}. The sources of
attenuation---such as these---that are clustered with matter will be
correlated with galaxies and large-scale structure, and can be found
with ``angular difference'' measurements that compare the apparent
properties of sources whose lines of sight have different impact
parameters with the correlated structure.

The Sloan Digital Sky Survey has permitted very sensitive angular
difference measurements, which find that the attenuation correlated
with large-scale structure is very small and
consistent with being caused by dust,
presumably the dust emitted by the stars in the galaxies that populate
the structure.  Measurements in the literature constrain this in
visible bandpasses at the part in $10^3$ level \citep{Menard2008,Bovy2008}.
To be specific, these studies constrain \emph{differences} in opacity
along different lines of sight caused by absorbers correlated with
galaxies.

It is possible, however, that there might be unclustered or
``monopole'' sources of attenuation, that affect all lines of sight
equally, for example if the nonmatter contributors to the
cosmological energy$-$momentum tensor (the ``dark energy'' in modern
parlance) have interactions with photons, or if there are small
violations of Lorentz invariance on cosmological scales.  These
sources of attenuation are much harder to detect with differential
experiments, but they \emph{can} be detected by ``radial difference''
experiments that compare cosmological sources of radiation of known
physical properties at different redshifts or radial distances.

A number of different mechanisms have been proposed during the last
decade to explain the observed dimming of type Ia supernovae (SNeIa;
\citealt{1998AJ....116.1009R,1999ApJ...517..565P}) without cosmic
acceleration by employing exactly such unclustered sources of
attenuation. The mixing of photons with axions in extragalactic
magnetic fields could lead to photons oscillating into axions with a
non-negligable probability over cosmological distances, thus reducing
the flux of SNeIa at large distances \citep{Csaki2002}. Alternatively,
``gray'' intergalactic dust could be so gray as to evade detection
through its \emph{reddening}, while still being cosmologically important
because of its overall \emph{opacity} \citep{1999ApJ...525..583A}. In order to
account for the observed SNeIa dimming, these models predict violations
of transparency at the order-unity level out to redshifts of unity
\citep[\eg, ][]{2002PhRvD..66d7702M}.

Furthermore, even if there are no exotic absorbers in the universe, it
is difficult (and usually model dependent) to infer the total mean
opacity from any absorbers that have been found by angular difference
experiments.  Radial difference and angular difference experiments are
complementary, even when the absorbers are mundane; although radial
difference experiments are usually less precise, they provide
irreplaceable information for measuring total opacity.

Radial difference experiments are sometimes known as ``Tolman tests''
because they are variants of the test proposed by \citet{Tolman1930}
of the expansion of the universe: a test of the $(1+z)^{-4}$ (where
$z$ is the redshift) dimming of bolometric intensity (energy per unit time
per unit area per unit solid angle; also called ``bolometric surface
brightness'') with redshift.  The intensity is closely related to the
phase-space density of photons, which is conserved (in a transparent
medium) along the light path; that conservation plus Lorentz
invariance implies the $(1+z)^{-4}$ relation: one factor of
$(1+z)^{-1}$ comes from the decrease in energy of each photon due to
the redshift, another factor comes from the decrease in photons per
unit time, and two more factors arise from the solid-angle effects of
relativistic aberration. The Tolman test does not really test for the
expansion of the universe---the result does not depend on cosmological
model, or even assumptions of isotropy or homogeneity---but
rather for the combination of conservation of photon
phase-space density and Lorentz invariance.

In addition to models that violate transparency, there are models that violate Lorentz invariance.  Generically
these models produce an energy-dependent speed of
light and birefringence, breaking the perfect nondispersiveness
of the vacuum \citep{1998Natur.393..763A,1999PhRvD..59l4021G}. These
effects generally become larger with increasing energy, and
observations of high-energy sources such as active galactic nuclei
\citep{1999PhRvL..83.2108B,2008arXiv0810.3475H,2008PhLB..668..253M} and gamma-ray bursts
\citep[GRBs;][]{1999PhRvL..82.4964S,2006APh....25..402E} have shown that the
linear dispersion relation for photons is preserved to good accuracy
at these energies. Therefore, while these models do fail the Tolman
test because of their nontrivial dispersion relations, the effect
will be unmeasurably small for low-energy (visible-band) photons.

By far the most precise radial difference test to date has been performed in the
radio with the cosmic background radiation.  In contemporary cosmological
models, the CBR comes from redshift $\sim 1100$ and is a near-perfect blackbody.
The COBE DMR experiment established that the spectrum and amplitude of this
radiation are consistent with the blackbody expectation at the $<10^{-2}$ level
at 95\% confidence \citep{Mather1994}.  A source of attenuating material,
unless in perfect thermal equilibrium with the CBR, would tend to change either
the spectrum or the amplitude, so this result provides a very strong constraint
on the transparency at cm wavelengths \citep{2005PhRvD..72b3501M}. Another test
of transparency at cm wavelengths is the increase in the CMB temperature $T_{\rm
CMB}$ according to the relation $T_{\rm CMB} \propto (1+z)$.
\citet{2000Natur.408..931W} find consistency with a transparent universe by
measuring the CMB temperature at $z=2.3$.  Of course, many sources of
attenuation are expected to be wavelength dependent, so these beautiful results
may not strongly constrain the opacity in the visible.

At visible wavelengths, there have been much less precise radial
difference tests performed with galaxies, whose properties would
deviate from naive predictions under extreme attenuation.  After
correcting for the evolution of stellar populations in galaxies, these
studies find consistency with transparency at the $0.5$ mag level at
95\% confidence \citep{pahre96a, lubin01a}, which correspond to
optical depth limits $<0.5$ out to redshift $z\sim 1$.  Unfortunately,
the precision of these tests is not limited by the precision of the
measurements, but rather by the precision with which the evolution of
galaxy stellar populations is known; the results will not be improved
substantially with additional or more precise observations.

Another test of transparency at visible wavelengths involves the
measurement of the Cosmic Infrared Background (hereafter CIB). The
absorption of visible photons by a diffuse component of intergalactic
dust and its re-emission in the infrared contributes to the CIB. The
amount of dust required to explain the systematic dimming of SNeIa
would produce most of the observed CIB
\citep{2000ApJ...532...28A}. However, discrete sources (\eg, dusty
star-forming galaxies) also emit in the infrared and account for
almost all of the CIB, strongly constraining the role of dust in the
dimming of SNeIa. Any constraint on the transparency from the CIB
requires a careful subtraction of the discrete sources
\citep{2001ARA&A..39..249H}.

The Tolman test can be rewritten as a relationship among cosmological
distance measures.  There are several empirical definitions of
distance in cosmology (\eg, \citealt{hogg99a}); the most
important for contemporary observables are the luminosity distance
$\DL$ and the angular diameter distance $\DA$.  The luminosity
distance $\DL$ to an object is defined to be the distance that relates
bolometric energy per unit time per unit area $S$ (flux) received at a
telescope to the energy per unit time $L$ (luminosity or power) of the
source, or
\begin{equation}
S = \frac{L}{4\pi\,\DL^2}
\quad .
\end{equation}
The angular diameter distance $\DA$ is the distance that relates the
observed (small) angular size $\Theta$ measured by a telescope to the
proper size $R$ of an object, or
\begin{equation}
\Theta = \frac{R}{\DA}
\quad .
\end{equation}
Because the ratio of flux to the solid angle is essentially the
intensity, the $(1+z)^{-4}$ redshift dependence of the intensity is
reflected in these distance measures by
\begin{equation}\label{DLDA}
\DL = (1+z)^2\,\DA
\quad .
\end{equation}
Both distance measures are strong functions of the world model, but
this relationship between them---known sometimes as the ``Etherington
relation'' (after \citealt{Etherington1933}, who showed that the
result is valid in arbitrary spacetimes)---depends only on conservation 
of phase-space
density of photons (transparency) and Lorentz invariance.
Fortunately, for some fortuitous types of objects, these distances can
be measured nearly independently.

A test of this type for transparency has been proposed and carried out
previously \citep{Bassett2004a, Bassett2004b}, with luminosity
distances from SNeIa and angular
diameter distances estimated from FRIIb radio galaxies, compact radio
sources, and X-ray clusters \citep{2004PhRvD..70h3533U,2008MNRAS.390L...1J}.
The results were imprecise because there
are many astrophysical uncertainties in the proper diameter estimates
of these exceedingly complex astrophysical sources.

In the contemporary adiabatic cosmological standard model, there is a
feature in the dark-matter autocorrelation function (or the power
spectrum) corresponding to the communication of density perturbations
by acoustic modes during the period in which radiation dominates
\citep{1970ApJ...162..815P,2005ApJ...633..560E}. This feature has a
low amplitude in present-day structure (that is, the distribution of
galaxies), but because it evolves little in comoving coordinates, it
provides a ``standard ruler'' for direct measurement of the expansion
history.  A measurement of the baryon acoustic feature (BAF) in a
population of galaxies at a particular redshift provides a combined measure
of the angular diameter distance to that redshift (from the transverse
size of the feature) and the Hubble constant or expansion rate at that
redshift (from the line-of-sight size of the feature).  As we discuss
below, as signal-to-noise improves, the BAF can be used to measure the
angular diameter independently of the local Hubble rate.  Most
importantly, because the BAF arises from extremely simple physics in
the early universe when the growth of structure is linear and
electromagnetic interactions dominate, the BAF measures the angular
diameter distance with far fewer assumptions than any method based on
complex astrophysical sources in the highly nonlinear regime.

At the same time, SNeIa have been found to be
standard---or really ``standardizable''---candles, which can be used
to make an independent direct measurement of the expansion history
\citep{1938ApJ....88..285B,Tammann1979,1979ApJ...232..404C,1998AJ....116.1009R,1999ApJ...517..565P}.
Up to an
overall scale and some uncertainties about the intrinsic spectra and
variability among SNeIa, a collection of SNeIa measure the luminosity
distance.

Given overall scale uncertainties, the most robust test of global
cosmic transparency that can be constructed from these two distance
indicators is a measurement of the ratio of the distances to two
redshifts $z_1$ and $z_2$.  That is, transparency requires
\begin{equation}
\frac{\DL(z_2)}{\DL(z_1)} = \frac{[1+z_2]^2}{[1+z_1]^2}\,
\frac{\DA(z_2)}{\DA(z_1)}
\quad .
\end{equation}
This expression cancels out overall scale issues and is independent of
the world model.  We perform a very conservative variant of this test
below, where we measure the left-hand side with SNeIa and the
right-hand side with the BAF, marginalizing over a broad range of
world models.


The tests presented here are not precise, simply because at the
present day BAF measurements are in their infancy, and we make use of
no cosmological data other than the BAF and SNeIa.  As we discuss
below, when these measurements are made at higher redshifts and with
higher precisions, our limits on transparency and Lorentz invariance
will improve by orders of magnitude.  Eventually they may be limited
not by the data quality but by the cosmic variance limit on the BAF
measurement itself \citep{Seo2007}.

\section{Data, procedure, and results}
\label{sec:data_analysis}

In surveys to date, where the BAF is measured at low signal to noise,
the optimal extraction of the signal best constrains not the angular diameter
distance directly, but rather a hybrid distance $\DV$
\begin{equation}
\DV = \left[\frac{c\,z\,[1+z]^2\,\DA^2}{H(z)}\right]^{1/3}
\quad ,
\end{equation}
where $\DA$ is the angular diameter distance and $H(z)$ is the Hubble
expansion rate (velocity per unit distance) at redshift $z$
\citep{2005ApJ...633..560E}.

Using data from the Sloan Digital Sky Survey and the Two Degree Field
Galaxy Redshift Survey, the power spectrum and BAF have now been
measured in samples of massive, red galaxies at two different
redshifts: $z=0.20$ and $z=0.35$. The measured BAF at each redshift
$z$ translates to a distance measure $\DV(z)$.  Accounting for
covariances in the measurements at the two redshifts (which are not
based on entirely independent data sets), the ratio of distances is
$\DV(0.35)/\DV(0.20)=1.812 \pm 0.060$ (68\% confidence;
\citealt{Percival2007}).

\begin{figure}
\resizebox{\textwidth}{!}{\includegraphics{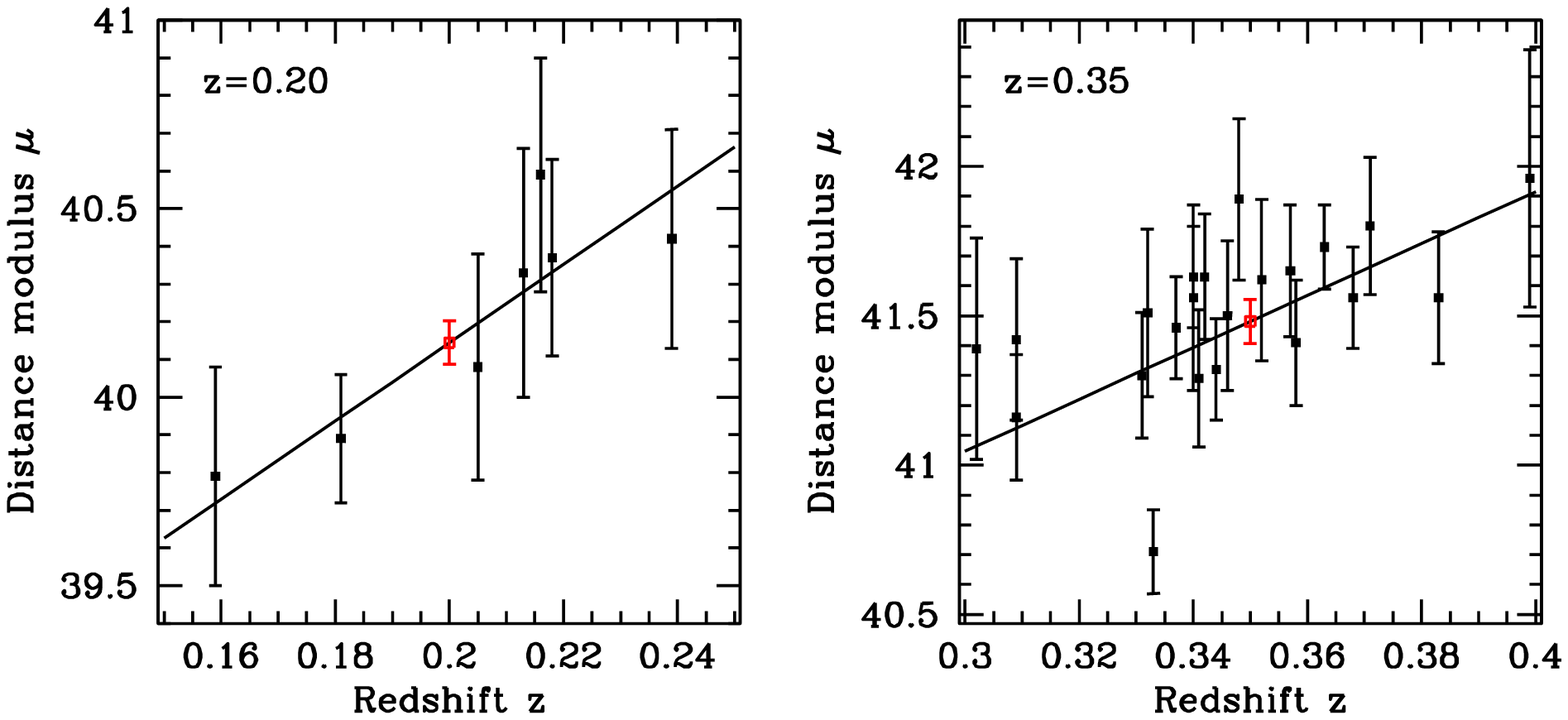}}
\caption{Distance-modulus--redshift relation.  Filled black
  squares with uncertainty bars show the SNeIa data (from
  \citealt{Davis2007}) used in Samples~A (left panel) and B (right
  panel).  Open red squares show the distance moduli $\DM(0.20)=
  40.14 \pm 0.06$ and $\DM(0.35) = 41.48 \pm 0.07$ (68\% confidence) 
  inferred from the fits to the data. \label{fig:fig1}}
\end{figure}

We formed two samples of SNeIa data from a recent compilation
\citep{Davis2007}.  ``Sample~A'' consists of all seven SNeIa in the
redshift range $0.15<z<0.25$ and ``Sample~B'' consists of all
22~SNeIa in the redshift range $0.3<z<0.4$.  We estimate the
distance$-$modulus $\DM$ at $z=0.20$ and $z=0.35$ by fitting a straight
line to Samples~A and B separately (\figurename~\ref{fig:fig1}), and
obtain a distance$-$modulus difference
\begin{equation}
\Delta\obs{\DM} = \obs{\DM}(0.35)-\obs{\DM}(0.20) = [1.34 \pm 0.09]~\mag
\quad ,
\end{equation}
where we are indicating that this is an observed value, and might differ
from the true value if there is opacity.

\begin{figure}
\resizebox{\textwidth}{!}{\includegraphics{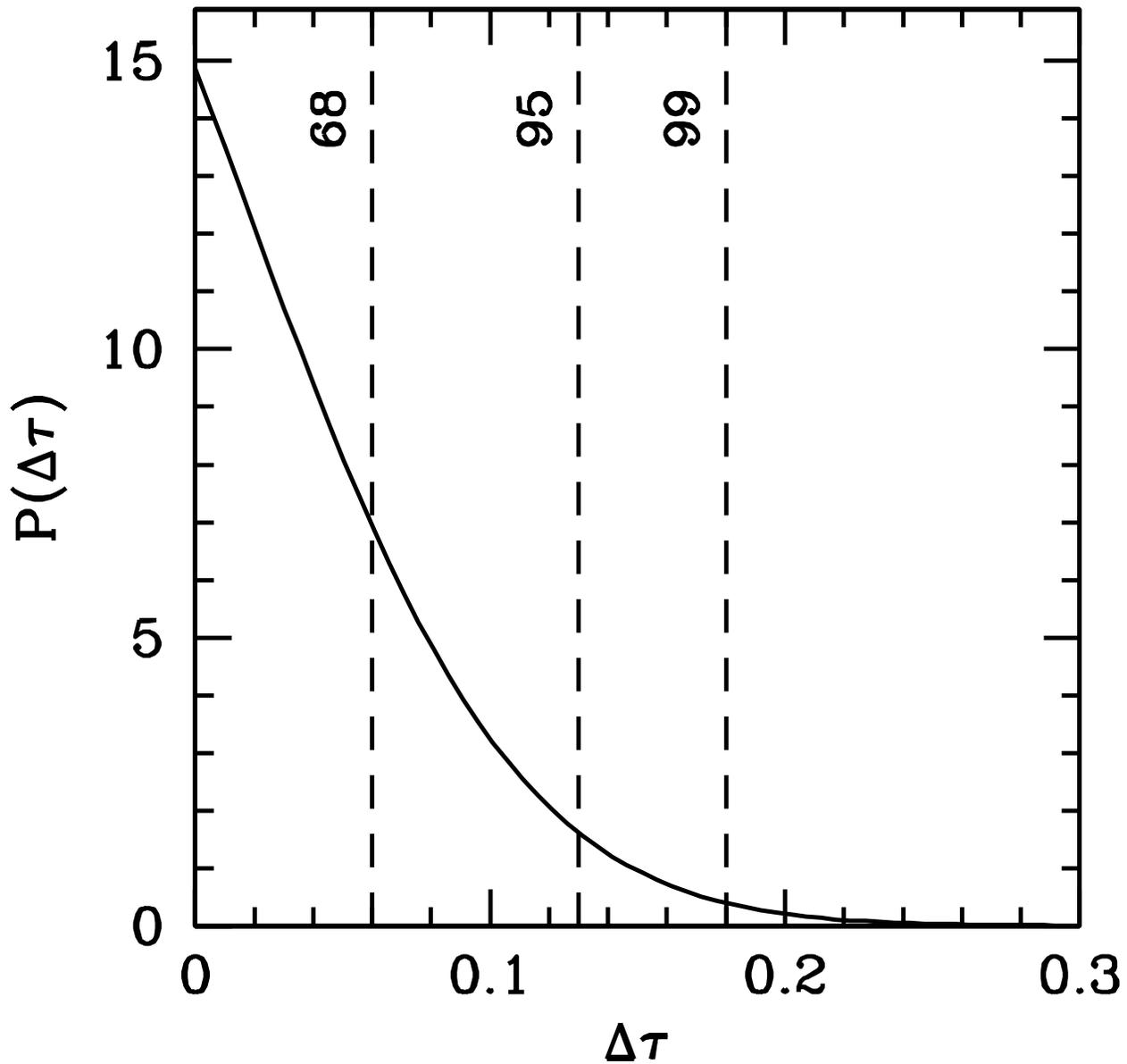}}
\caption{Posterior distribution of $\Delta \tau$ between $z=0.35$ and
  $z=0.20$ obtained from the Bayesian analysis described in Section~
  \ref{sec:data_analysis}. The 68, 95 and 99\% confidence upper
  limits are indicated by the corresponding dashed lines.
\label{fig:fig2}}
\end{figure}

The distance modulus derived from the SNeIa is systematically affected
by the presence of any intervening absorber. Let $\tau(z)$ denote the 
opacity between an observer at $z=0$ and a source at redshift $z$ due to such 
extinction effects. The flux received from this source is reduced by the 
factor $\exp{-\tau(z)}$. The inferred (observed) luminosity distance 
differs from the ``true'' luminosity distance:
\begin{equation}
\obs{\DL}^2(z) = \true{\DL}^2(z)\,\exp{\tau(z)}
\quad .
\end{equation}
The ratio of the luminosity distances at two different redshifts $z_1$ and $z_2$ 
depends upon the factor $\exp{[\tau(0.35)-\tau(0.20)]/2}$. The inferred 
(observed) distance modulus differs from the ``true'' distance modulus:
\begin{equation}
\obs{\DM}(z) = \true{\DM}(z)+[2.5\,\log e]\,\tau(z)
\quad .
\end{equation}
Taking differences of distance moduli at the two redshifts:
\begin{equation}
 \obs{\Delta\DM} = \true{\Delta\DM}+[2.5\,\log e]\,\Delta\tau
\quad ,
\end{equation}
where $\Delta\tau\equiv[\tau(z_2)-\tau(z_1)]$.  If the distance
indicator from the BAF is unaffected by the absorption as we expect,
then
\begin{equation}
\Delta\tau
 =\frac{\ln(10)}{2.5}\,\left[\obs{\Delta\DM}
 -7.5\,\log\left(\frac{\DV(z_2)}{\DV(z_1)}\right)
 +2.5\,\log\left(\frac{z_2\,[1+z_1]^2\,H(z_1)}
                      {z_1\,[1+z_2]^2\,H(z_2)}\right)\right]
\quad .
\label{dtaumod}
\end{equation}
The above equation can be used to determine $\Delta\tau$ from $z=0.35$
to $z=0.20$ in light of the ratio of the distances $\DV$ obtained from
the BAF observations (hereafter $B$) and the difference in distance
moduli obtained from the SNeIa observations (hereafter $S$) at these
redshifts.  However, the last term in the above equation makes the
result cosmology dependent. Therefore, we follow a Bayesian approach
and assign posterior probabilities to 100 uniformly spaced values of
$\Delta\tau \in $ [0,0.5] by marginalizing over $100\times100$
$\Lambda$CDM cosmologies uniformly spaced in the
$(\Omega_\Lambda,\Omega_M)$ plane with $\Omega_\Lambda \in $ [0,1] and
$\Omega_M \in $ [0,1]. Thus,
\begin{equation}
P(\Delta\tau|S,B) = \int_{\Omega_\Lambda} \int_{\Omega_M}
  \,P(\Omega_\Lambda,\Omega_M|B)
  \,P(\Delta\tau,\Omega_\Lambda,\Omega_M|S)
  \,\drm\Omega_M\,\drm\Omega_\Lambda
\quad ,
\label{marginalise}
\end{equation}
where $P(\Omega_\Lambda,\Omega_M|B)$ and
$P(\Delta\tau,\Omega_\Lambda,\Omega_M|S)$ are the posterior
probabilities of the set of model parameters given $B$ and $S$
respectively. We assume that the uncertainties on $B$ and $S$ are Gaussian
and calculate the likelihood of $B$ and $S$ for different sets of
parameters in the ($\Delta\tau,\Omega_\Lambda,\Omega_M$) space.
Assuming flat priors on $\Omega_\Lambda$ and $\Omega_M$ in the ranges
$0<\Omega<1$, and flat prior on $\Delta\tau$ in the range
$0<\Delta\tau<0.5$, the posterior probabilities
$P(\Omega_\Lambda,\Omega_M|B)$ and
$P(\Delta\tau,\Omega_\Lambda,\Omega_M|S)$ are calculated from the
likelihoods of the two data sets. \equationname~(\ref{marginalise})
yields the posterior for $\Delta\tau$, marginalized over all world
models. \figurename~\ref{fig:fig2} shows the posterior
$P(\Delta\tau|S,B)$ for the difference in optical depths between
redshifts $0.35$ and $0.20$ obtained from the procedure outlined
above. The posterior peaks at $0$ and yields $\Delta\tau<0.13$ at
95\% confidence. The result demonstrates the transparency of the
universe between these two redshifts, although not at high precision.

The abundance and absorption properties of absorbers can be constrained 
using the difference in optical depths measured above. Let $n(z)$ denote the
comoving number density of absorbers, each with a proper cross-section 
$\sigma(z)$ at redshift $z$. The difference in optical depths between
redshifts $z_1$ and $z_2$ is then given by
\begin{equation}
\Delta\tau = \int_{z_1}^{z_2} n(z)\,\sigma(z)\,\DHub\,\frac{(1+z)^2}{E(z)}
  \,\drm z
\quad ,
\end{equation}
where $\DHub$ is $c/H_0$ and
\begin{equation}
E(z) \equiv \frac{H(z)}{H_0}
  = \sqrt{\Omega_M(1+z)^3+\Omega_k(1+z)^2+\Omega_\Lambda}
\quad .
\end{equation}
In detail, the output of this integral depends on world model.  For
the concordance model,
Hubble constant $H_0=100\,h~\km\,\s^{-1}\,\Mpc^{-1}$,
and assuming $n(z)$ and $\sigma(z)$ to be
independent of redshift, $\Delta\tau$ measured between redshifts 0.35
and 0.20 constrains $n\,\sigma<2\times10^{-4}\,h~\Mpc^{-1}$ at
95\% confidence.

A naive calculation of $\Delta\tau$ using
\equationname~(\ref{dtaumod}) for the concordance $\Lambda$CDM model
($\Omega_M=0.258,\Omega_\Lambda=0.742$) obtained from the analysis of
the 5-year WMAP data \citep{Dunkley2008}, yields $\Delta\tau =
-0.30\pm0.26$ at 95\% confidence. This shows that there is a
slight tension between the results of current measurements of the BAF
and of the SNeIa under the currently accepted world model. More
generally, a similar tension, \ie, a brightening of the SNe, between
measurements of the cosmological parameters by using standard rulers
and standard candles has been reported before
\citep{Bassett2004a,Bassett2004b,Percival2007,2008JCAP...07..012L}.
SNe brightening is not impossible in models that involve axion-photon
mixing \citep{Bassett2004b} or chameleon-photon mixing
\citep{2008PhRvD..77d3009B} if the corresponding particles are
abundantly produced during SNeIa explosions. However, a negative value
of $\Delta\tau$ could also indicate the presence of a systematic bias
in the distance measurements based upon the SNeIa brightness or the
BAF, \eg, overcorrection for extinction in the host galaxy of the
SNeIa brightnesses or magnification bias in the SNeIa selection
\citep{2004MNRAS.351.1387W}. Note that the prior, $\Delta\tau > 0$,
improves the magnitude of the uncertainty on $\Delta\tau$ (from 0.26
to 0.13).  The 95\% confidence interval shrinks with the prior
because we sample only from the rapidly falling tail of the posterior.

\section{Future Constraints}

In the future, the constraints from both the SNeIa and the BAF
observations will improve in accuracy and will cover a wider redshift
range. The Baryon Oscillation Spectroscopic Survey (BOSS) is currently
underway and plans to measure the BAF in luminous red galaxies at
redshifts $z=0.35$ and $z=0.6$. The key improvements would be the
larger redshift range and the power to resolve the BAF both in the
line-of-sight direction (constrains $H$) and the transverse direction
(constrains $\DA$).  This would remove the weak world-model dependence
in our present analysis.  The angular diameter distances to these
redshifts would be measured to an accuracy of $\sim 1$\%
(http://www.sdss3.org/). In parallel, the Supernovae Legacy Survey
(SNLS), when complete, expects $\sim 700$ SNeIa in the redshift range
$0<z<1.7$ \citep{2006A&A...447...31A}. The uncertainty on the estimate
of the distance moduli to redshifts $z=0.35$ and $z=0.6$ will be
roughly four times better with the increased numbers.  Using the test
of the duality relation described above, $\Delta \tau$ between
$z=0.35$ and $z=0.6$ would be constrained to better than $0.07$
(95\% confidence), independent of the adopted cosmological
model. The constraint on $n\,\sigma$ would become
$n\,\sigma<5.4\times10^{-5}\,h~\Mpc^{-1}$.

BOSS will also use the Ly$\alpha$ forest in the spectra of bright
quasi-stellar objects to measure the BAF at redshift $z\sim2.5$ with
an accuracy of $\sim 1.5$~percent. No current or planned SNeIa surveys
expect to detect SNeIa at such a high redshift. However the highest
redshift ($\sim 1.7$) measurements of $\DL$ from the SNLS could
potentially be used in conjunction with the $\DA$ measurement to get a
constraint on the transparency of the universe by marginalizing over
different world models. Interestingly, there have been recent efforts
to calibrate GRBs as standard candles and
to extend the Hubble diagram to higher redshifts
\citep{2008ApJ...685..354L}. The SNeIa at low redshift and the GRBs at
high redshifts can provide a measurement of the difference between the
$\DM$ between redshifts $0.35$ and $2.5$. We optimistically assume
that the difference in the DM to these redshifts can be measured with
an accuracy of $\sim0.1$ similar to the one obtained from the analysis
of SNeIa at $z=0.2$ and $z=0.35$ in Section~\ref{sec:data_analysis}.
These measurements shall then constrain $\Delta \tau$ between
redshifts $2.5$ and $0.35$ to an accuracy of $0.2$ with $95$~percent
confidence. This translates into an accuracy on $n\,\sigma$ of
$\sim1.1\times10^{-5}\,h~\Mpc^{-1}$.

In the optimistic future, the uncertainty on $\DL(z)$ could, in the
absence of damaging systematics, diminish arbitrarily as the number of
observed SNeIa grows. However, the precision of any BAF measurement is
limited by sample variance (the number of independent wavelengths of a
given fluctuation that can fit in the finite survey volume is
limited), even when the uncertainty caused by incomplete sampling of
the density field (shot noise) is negligible \citep{Seo2007}. The
sample variance error goes down with the square root of the volume of
the survey.  To calculate a representative limit, we consider an
optimistic all-sky survey covering the redshift range
$2.45<z<2.55$. Such a survey can be used to determine $\DA(z=2.5)$ to
a fractional accuracy of $\sim 0.004$ (95\% confidence).  This
will ultimately constrain the optical depth to redshift $z=2.5$ to
$\tau<0.008$ and hypothetical absorbers to
$n\,\sigma<4\times10^{-7}\,h~\Mpc^{-1}$.

\section{Discussion}

We have advocated and analyzed the expected future performance of a
simple Tolman test or test of the Etherington relation, that is, that
the luminosity distance is larger than the angular diameter distance
by two powers $(1+z)$, using SNeIa to measure the
luminosity distance and the BAF to measure the
angular diameter distance.  We have shown that this test will
eventually provide very precise measurements of the conservation of
photon phase-space density.

We performed the test with the limited data available at the present
day.  We used only the ratio of distances at redshifts of $z=0.20$ and
$0.35$ to remove uncertainties about the overall scale.  We find
consistency with a Lorentz-invariant, transparent universe.  Our
results are consistent with all other measures of transparency to
date.  This is in part because they are not extremely precise.  Our
Tolman test also assumes that the measurements of the SNeIa and of the
BAF are not affected by systematic biases with magnitudes that are a
significant fraction of the magnitudes of the uncertainties.  Our test
is limited by the precision of the BAF measurement and the redshift
range over which it has been measured.  As we have shown, experiments
planned and underway will increase the redshift range and improve the
overall precision by an order of magnitude.

The most precise transparency measurements at visible wavelengths today
are statistical angular difference measurements, which can only
constrain attenuation correlated with specific types of absorbing
structures in the universe (\eg, MgII absorbers, \citealt{Menard2008};
clusters of galaxies, \citealt{Bovy2008}).  The simple Tolman
test performed here limits the full, unclustered, line-of-sight
attenuation between two redshifts.

The technique used in this paper provides a test of transparency that
is not very sensitive to astrophysical assumptions, both because the
BAF has a straightforward origin during an epoch in which growth of
structure is linear and the dominant physics is well understood, and
because there is no significant ``evolution'' with cosmic time for
which we must account.  This is in contrast to other methods for
measuring angular diameter distances and brightnesses, where there are
no precisely ``standard'' rulers, and evolution is dramatic with
redshift.  On the other hand, the ultimate precision of any test of
this type may come from the finite comoving volume in the observable
universe.  Cosmic variance will dominate the BAF error budget
eventually.

The SNeIa samples have been corrected as best as possible for
line-of-sight extinction by fitting an empirical correlation of
extinction with a change in color.  However, there are a few problems
with this approach.  First, this approach cannot correct for ``gray''
dust \citep{1999ApJ...525..583A}.  Second, this approach can also not
correct for a monopole component; it only corrects for components that
show variations around the mean level.  Third, these corrections will
be wrong or fooled if there are intrinsic relationships between color
and luminosity for SNeIa.  Fourth, the empirical corrections found by
these projects tend to be odd in the context of what is expected from
the reddening and attenuation by dust in the Milky Way
\citep{1982Ap.....18..328J,2007ApJ...664L..13C,2008ApJ...674...51E,
2008A&A...487...19N}.
The Tolman
test is sensitive to any kind of absorber and makes no assumptions
about the wavelength dependence or fluctuations of the opacity.  Given
that the SNeIa results have been corrected for a color--brightness
relation, the test presented here looks at the mean opacity toward
SNeIa of the fiducial color to which the compiled SNeIa have been
corrected.

The best-fit value of $\Delta\tau$ obtained from our analysis is
negative, \ie, SNeIa are brighter than expected from the angular
diameter distance measurements using the BAF. A conversion of dark
sector particles into photons could provide a physical explanation for
this result. However, this could also indicate the presence of a
systematic bias in either the SNeIa or BAF experiments. A test such as
the one presented in this paper is a useful tool to bring such biases
to light.

At present, because the differences among competitive world models are
not large over the redshift range $0.20<z<0.35$, our test is not yet
sensitive enough to rule out extreme axion or ``gray'' dust models
that reconcile SNeIa results with an Einstein-de~Sitter universe by
using effective opacity to adjust the inferred
redshift--luminosity-distance relation.  However, these models will
all be severely constrained within the next few years \citep[see
also][]{2006MNRAS.372..191C}.

\acknowledgements It is a pleasure to thank Frank C.~van den Bosch,
Aaron Chou, David Cinabro, Daniel Eisenstein, Zoltan Haiman, Engelbert
Schucking and Michael Strauss for comments and assistance.  S.M. is a
member of the IMPRS for Astronomy \& Cosmic Physics at the University
of Heidelberg. D.W.H. was partially supported by NASA (grant
NNX08AJ48G) and a research fellowship of the Alexander von Humboldt
Foundation of Germany.

\clearpage

\end{document}